\documentclass[journal]{IEEEtran}

\usepackage{xcolor,soul,framed} 

\colorlet{shadecolor}{yellow}
\usepackage[pdftex]{graphicx}
\graphicspath{{../pdf/}{../jpeg/}}
\DeclareGraphicsExtensions{.pdf,.jpeg,.png}

\usepackage[cmex10]{amsmath}
\usepackage{array}
\usepackage{mdwmath}
\usepackage{mdwtab}
\usepackage{eqparbox}
\usepackage{url}

\usepackage{blindtext}
\usepackage{multirow}
\usepackage{mathtools} 
\usepackage[noadjust]{cite}
\usepackage{float}
\usepackage[letterpaper, margin=0.63in,top= 0.725 in,right=0.64in, left=0.64in]{geometry}
\usepackage{bbm}
\usepackage{amsthm}
\usepackage{cite}
\usepackage{amsfonts}
\usepackage{algorithm}
\usepackage{algpseudocode}
\usepackage{amsmath}
\usepackage[utf8x]{inputenc}
\usepackage{subcaption}
\usepackage{placeins}

\newtheorem{definition}{Definition}

\algnewcommand\algorithmicforeach{\textbf{for each}}
\algdef{S}[FOR]{ForEach}[1]{\algorithmicforeach\ #1\ \algorithmicdo}

\usepackage{hyperref} 
\usepackage{orcidlink}
\begin{document}
    \title{A$^3$L-FEC: Age-Aware Application Layer Forward Error Correction Flow Control}
  \author{
       \IEEEauthorblockN{Sajjad~Baghaee\,\orcidlink{0000-0003-2156-3403},~\IEEEmembership{Member,~IEEE,}~Elif~Uysal\,\orcidlink{0000-0002-7258-4872},~\IEEEmembership{Fellow,~IEEE}}
  \thanks{Manuscript received .. .., 2025; revised .. .. 2025; accepted .. .. 2025; approved by ..... This work was funded by Scientific and Technological Research Council of Turkey (TUBITAK) under the Grant Number 22AG019.}
  \thanks{S. Baghaee and E. Uysal are with Department of Electrical and Electronics Engineering, Middle East Technical University, 06800 Ankara, Turkey (e-mail: sajjad@baghaee.com, uelif@metu.edu.tr).}}

\markboth{(Under review)}{Baghaee \MakeLowercase{\textit{et al.}}: A$^3$L-FEC: Age-Aware Application Layer Forward Error Correction Flow Control}

\maketitle

\begin{abstract}
\label{abstract}
Age of Information (AoI) is a metric and KPI that has been developed for measuring and controlling data freshness. Optimization of AoI in a real-life network requires adapting the rate and timing of transmissions to varying network conditions. The vast majority of previous research on the control of AoI has been 
theoretical, using idealized models that ignored certain implementation aspects. As such, there is still a gap between the research on AoI and real-world protocols. In this paper we present an effort toward closing this gap by introducing an age-aware flow control algorithm. The algorithm, Age-Aware Application Layer Forward Error
Correction (A$^3$L-FEC), is a packet generation mechanism operating on top of the User Datagram
Protocol (UDP). The purpose is to control the peak Age of the end-to-end packet flow, specifically to reduce the rate of so-called “Age Violations,” i.e., events where the peak age exceeds a given threshold. Evaluations in Mininet-WiFi and MATLAB indicate that A$3$L-FEC reduces age violations compared to two related protocols in the literature, namely TCP-BBR and ACP+.
\end{abstract}
\IEEEpeerreviewmaketitle
\vspace{-0.6cm}

\section{Introduction}

\IEEEPARstart{T}{he} rapid growth of the Internet of Things (IoT) and applications such as remote monitoring and automation has led to a surge in time-sensitive data flows. These systems rely on the timely delivery of fresh information to ensure accurate decision-making and system reliability. To capture the freshness of received updates, the Age of Information (AoI) was introduced as a key performance indicator (KPI)~\cite{2011KaulGruteserRaiKenney}.

AoI is the time elapsed since the most recently received packet was generated at the source. Formally, at time~$t$, it is given by $\Delta(t) = t - t'$, where $t'$ is the generation time of the freshest received packet. Unlike traditional metrics such as delay or throughput, AoI captures both latency and update frequency, providing a measure of freshness over time. For example, in a First-Come-First-Served (FCFS) queue, infrequent updates yield high AoI despite low delay, while excessive updates may congest the network and again increase AoI. Thus, AoI control requires balancing update rate and congestion.

Recently, AoI has become increasingly relevant in the context of 5G and beyond, particularly for massive machine-type communications (mMTC)\cite{10634462} and Ultra-Reliable Low-Latency Communications (URLLC)\cite{10415249}. With the evolution toward eXtreme URLLC (xURLLC) in 6G, AoI and related metrics are expected to play a central role in supporting real-time services~\cite{10355071}. While average AoI is commonly used as a KPI, it may not sufficiently reflect the urgency of information in mission-critical (MC) systems~\cite{10663282}. In such scenarios, minimizing the age violation probability—the probability that AoI exceeds a predefined threshold—provides a more task-relevant measure to ensure timely and reliable updates.

Achieving low AoI in practice is challenging due to dynamic network conditions. As traffic load approaches or exceeds capacity, queuing delays and packet loss increase, degrading the timeliness of updates. Techniques such as adaptive sampling and priority scheduling can help, but they require a flow control mechanism that accounts for information freshness.


\textbf{\textit{Contribution:}} To address this, we propose A$^{3}$L-FEC (Age-Aware Application Layer Forward Error Correction), a practical flow control algorithm operating at the application layer. A$^{3}$L-FEC uses rateless forward error correction (FEC) over the User Datagram Protocol (UDP) to adapt redundancy and transmission rate according to network conditions and freshness requirements. By injecting just enough redundancy to maintain timely updates without causing congestion, it aligns with the principle of “\textit{keeping the pipe just full, but no fuller}”~\cite{Kleinrock2018InternetCC}.
\vspace{-0.2cm}
\section{Related Work}

Congestion control algorithms in the Internet can be broadly grouped into three categories based on their congestion signals and optimization goals~\cite{Turkovic2019FiftySO}: Loss-based algorithms, such as Reno and Cubic, use packet loss as a congestion signal. While designed to maximize throughput, they often induce excessive queuing and delay, especially over wireless bottlenecks. Delay-based algorithms, including Vegas and FAST, attempt to maintain low queuing delay by using round-trip time as the signal. However, they often misinterpret network delay due to noise, such as ACK compression and jitter, leading to underutilized links. To overcome these limitations, hybrid algorithms like BBR estimate both bottleneck bandwidth and round-trip delay. BBR aims to operate at the bandwidth-delay product (BDP), proactively minimizing delay while having high throughput.

However, these algorithms are not designed to optimize information freshness. By prioritizing throughput, they may fill queues with outdated packets—resulting in high AoI. This makes them inadequate for time-sensitive applications, where timely delivery is more critical. These limitations underscore the need for congestion controls that explicitly account for AoI to ensure fresh updates under varying network conditions.

The AoI metric has gained increasing attention as a measure of data freshness across various network models~\cite{survey2,Uysal_Kaya_Baghaee_Beytur_2023}, though relatively few studies have explored its optimization in real-world settings. Early emulation efforts include~\cite{core-aoi}, which evaluated AoI under diverse wireless scenarios using CORE and EMANE, and~\cite{canberk2018}, which demonstrated a non-monotonic “U-shaped” relationship between AoI and rate over TCP/IP.

In~\cite{HasanRealLife}, AoI was measured experimentally in a two-way connection between a transceiver and echo server, highlighting the impact of real-world factors such as propagation delay and queuing. Further,~\cite{HasanSmartIoT} compared AoI across TCP, UDP, and WebSocket on wired and wireless links testbed.

Reinforcement learning has also been applied to AoI control. In~\cite{EgemenRL}, the AoI optimization problem was formulated as a Markov Decision Process, enabling model-free rate adaptation over the Internet. In~\cite{10032102}, the authors studied AoI-aware scheduling in IIoT systems with both periodic and random sampling, introducing Lyapunov-based and D3QN policies to minimize AoI under delay constraints in noisy channels.

WiFresh~\cite{kadota2020wifresh} introduced an AoI-optimized WiFi scheduler, achieving up to two orders of magnitude improvement in freshness under heavy load. Similarly, WiSwarm~\cite{10228860} proposed an AoI-aware middleware for UAV teams, prioritizing timely updates and improving tracking performance.



The Age Control Protocol (ACP)\cite{ACP2018} and its enhanced version ACP+\cite{10483026} were developed to support timely updates over IP networks. While ACP+ performs well on fat pipe and long-path connections, its implementation on low-latency, short-haul IoT devices has proven challenging~\cite{ACPUmut}. Although ACP and ACP+ are closely related to our work, A$^3$L-FEC differs in operation. While ACP variants adjust transmission rates based solely on instantaneous age, A$^3$L-FEC actively prevents age violations by incorporating memory of past delay, age, and violations, enabling more adaptive and stable flow control.

Beyond rate control, AoI-aware coding schemes like~\cite{costa2020robust} use random linear coding under in-order delivery and lossless assumptions. In contrast, A$^3$L-FEC employs sample-based coding that tolerates individual losses and includes congestion control.


\section{System Model}
\label{a3lfecsystemModel}

AoI quantifies the time elapsed since the most recently received update was generated at the source. It serves as a measure of information freshness at the destination. A typical evolution of AoI over time follows a sawtooth pattern: it increases linearly between updates and drops upon the reception of a fresher packet. An age violation occurs when the instantaneous age, $\Delta(t)$, exceeds a predefined threshold $AVT$, indicating that the received data is outdated.


\begin{figure}
\centering
         \includegraphics[scale=0.35]{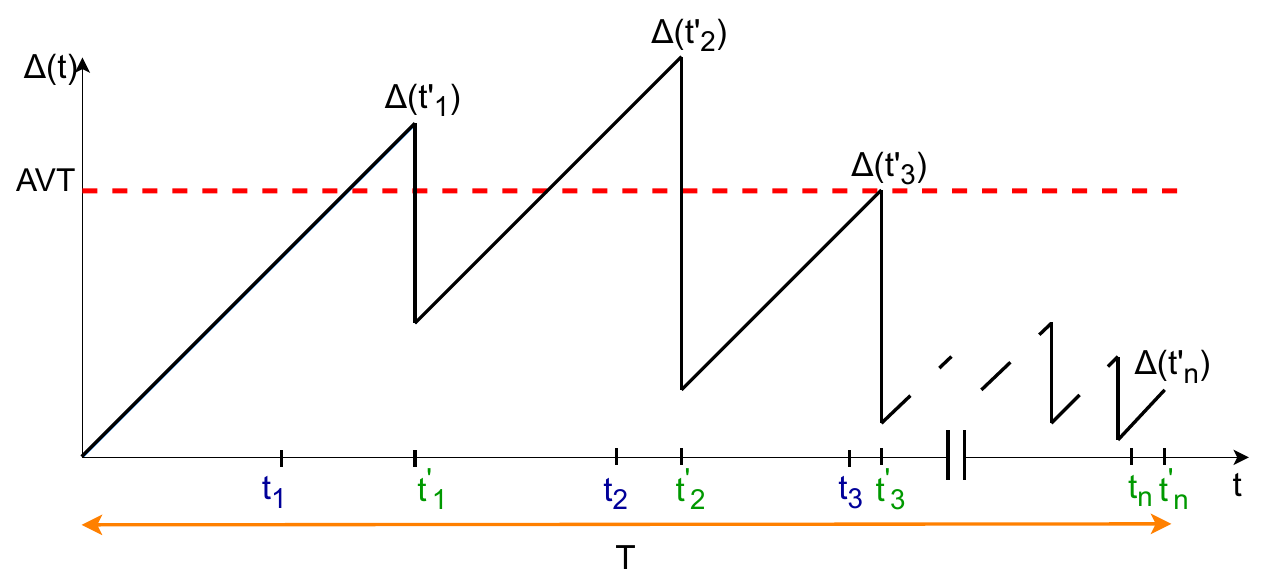}
\caption{Sample path of the age process $\Delta(t)$, \cite{10601159}.}
\label{fig:aoi}
\end{figure}


This study presents two variants of A$^{3}$L-FEC, each designed for different operating conditions. A$^{3}$L-FEC-FSFB targets fixed sampling rates and block lengths, while A$^{3}$L-FEC-VSVB supports variable sampling and block lengths. Both share the same core mechanism but differ in how updates are generated and transmitted. The following subsections detail these differences.
\subsection{\texorpdfstring{A$^3$L-FEC}{\space}-Fixed Sampling Rate Fixed
        Block-length}
\label{a3lfecsystemModelFSFB}

\begin{figure}[h]
        \centering
        \includegraphics[scale=0.26]{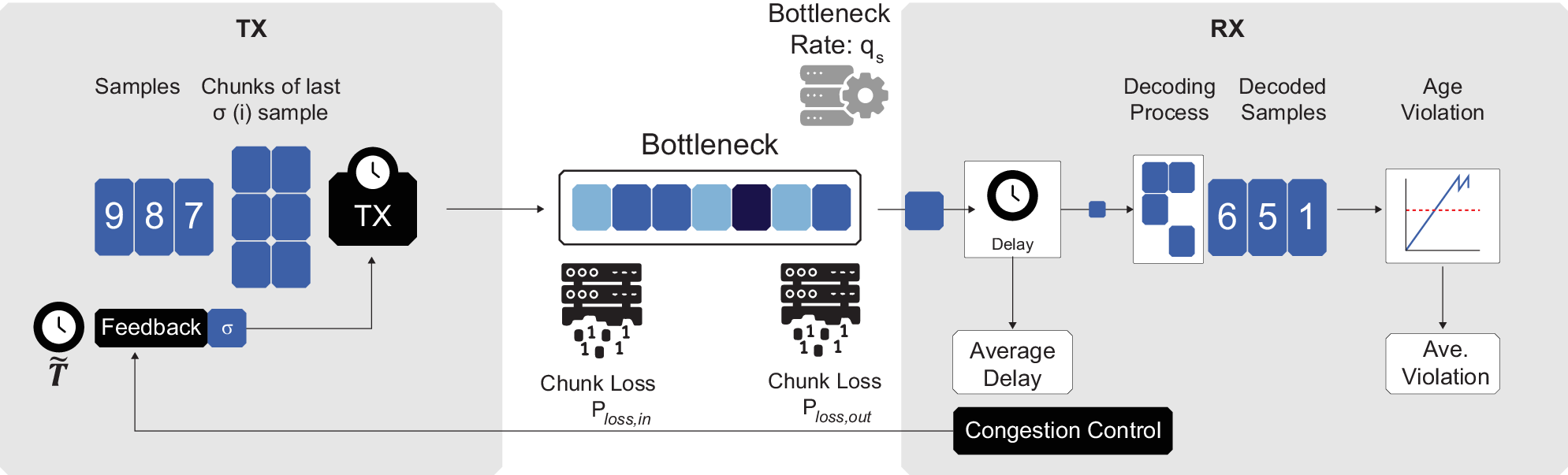}
        \caption{System model of a status update system over an error-prone link with a transmission rate feedback.}
        \label{fig:sms1}
\end{figure}

From an application-layer perspective, A$^3$L-FEC-FSFB operates in a time-slotted status update system over an error-prone link (Fig.~\ref{fig:sms1}). A receiver monitors a time-varying source process by receiving updates from a remote transmitter via a Packet Erasure Channel (PEC). The source, denoted by $S$, generates samples $\{ s_{\tau} \}_{\tau \geq 0}$ at fixed intervals. Throughout this study, all time durations are expressed in units of a single time slot.

\begin{figure}
        \centering
        \includegraphics[scale=0.70]{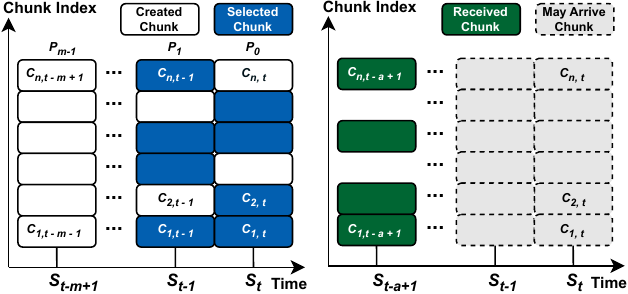}
        \caption{Chunks in the A$^3$L-FEC-FSFB protocol.}
        \label{fig:codewordsampleTX}
\end{figure}

At the transmitter, each sampled value of length $K$ bits is divided into $k$ equal-length chunks, then encoded into $n$ coded chunks of size $\frac{K}{k}$ bits using a Maximum Distance Separable (MDS) code. This allows the receiver to reconstruct the original sample from any $k$ out of $n$ coded chunks. Let $c_{i,\tau}$ denote the $i^{\text{th}}$ coded chunk of sample $s_\tau$, generated at time $\tau$. Chunks are transmitted via UDP, each packet carrying a chunk and its index. The transmitter may send multiple packets per time slot and buffers only chunks from the $m$ most recent samples. Let $\mathcal{T}_{t}$ be the set of chunks transmitted in slot $t$, where $c_{i,\tau} \in \mathcal{T}_{t}$ implies that chunk $i$ of sample $s_{\tau}$ is sent at time $t$. Thus, $\mathcal{T}_{t} \subset \{ c_{i,\tau}: i=1, \ldots, n~;~\tau=t-m+1, \ldots, t \}$
(Fig. \ref{fig:codewordsampleTX}).

The performance of A$^3$L-FEC-FSFB is evaluated using a First-Come-First-Served (FCFS) queuing model. The transmitter sends UDP packets at a rate of $\tilde{\sigma}$ [codewords/time slot] through a single infinite-buffer bottleneck with service rate $q_s$ [chunks/time slot]. Each chunk takes $\frac{1}{q_s}$ time slots to transmit, assuming no queuing delay. A packet may be dropped before entering the bottleneck with probability $P_{\text{loss,in}}$, or after, with probability $P_{\text{loss,out}}$. The total packet loss probability is thus:

\begin{align}
        \label{eq:Pcloss}
        P_{loss,c}= P_{loss,in} + (1-P_{loss,in})P_{loss,out}.
\end{align}



Due to the network bottleneck, each packet experiences a random delay before reaching the receiver. Let $D_{t;i,\tau}$ denote the delay of coded chunk $c_{i,\tau}$ sent at time $t$. If the packet is lost or corrupted, the delay is set to infinity: $D_{t;i,\tau} = \infty$.

The receiver stores successfully received chunks to reconstruct codewords. Let $\mathcal{C}_t$ be the set of coded chunks received by time $t$, and $\mathcal{S}_{t}$ the set of decodable samples. A sample $s_\tau$ is undecodable at $t$ if more than $n-k$ of its chunks are missing:

\vspace{-0.1cm}
\begin{equation}
        \label{indicatorfunctionformissingsample}
        {\{ s_{\tau} \notin \mathcal{S}_{t}\}}= {\{\sum_{i=1}^{n}\mathbbm {1}_{\{ c_{i,\tau} \notin \mathcal{C}_{t} \}}>n-k \}},
\end{equation}
where $\mathbbm{1}$ is the indicator of a missing coded chunk event:

\begin{equation}
        \label{indicatorfunctionformissingcodedchunks}
        \mathbbm{1}_{\{ c_{i,\tau}\notin \mathcal{C}_{t} \}}=\prod_{j=\tau}^{t}\left( \mathbbm{1}_{\{ c_{i,\tau}\notin \mathcal{T}_{j} \}}+ \mathbbm{1}_{\{ c_{i,\tau}\in \mathcal{T}_{j} \}}\mathbbm{1}_{\{ D_{t;i,\tau}> t-j \}}\right).
\end{equation}

In our system, the AoI at the receiver is defined as:

\begin{equation}
        \Delta(t)=t-\max \{ \tau \in \{ 1,2,...,t\} :  s_{\tau} \in \mathcal{S}_{t}\}.
\end{equation}


Given an age violation threshold $AVT$, the objective of the A$^3$L-FEC is to minimize the age violation rate, denoted by $AV$. As defined in~(\ref{eq:ageVoilationMetric1}), $AV$ quantifies the quality of monitoring the source process $S$ over a time window of duration $T$.

\begin{equation}
        \label{eq:ageVoilationMetric1}
        AV=\frac{1}{T}\displaystyle\sum_{t=1}^{T}\mathbbm{1}
        _{\{ \Delta(t) \geq AVT \}},
\end{equation}


To improve monitoring quality, we adopt a \emph{Stationary Independent Selection} (SIS) policy, which optimizes sample selection to consistently deliver fresh updates to the receiver.

\begin{definition}: A transmission policy is called a stationary independent
        selection (SIS) policy, if the events ${\{ c_{i,\tau}  \in \mathcal{T}_{t} \}}$
        are independent for all $i$, $\tau$ and $t$ such that ${\{ c_{i,\tau}
                                \in \mathcal{T}_{t} \}}$ occurs with probability $p_{t-\tau}$ where
        $p_{v}=0$ for $v < 0$ and $v > m-1$.
        \label{def1}
\end{definition}


Under Definition~\ref{def1}, sample selection follows stationarity and independence, meaning that events $c_{i,\tau} \in \mathcal{T}_{t}$ are independent across all $i$, $\tau$, and $t$. The policy is governed by a probability sequence $p_0, \ldots, p_{m-1}$ that defines the selection process.



In our threshold-based AoI framework, it is essential to avoid transmitting outdated samples. To enforce this, the transmitter filters out samples whose generation times exceed the age violation threshold $AVT$, ensuring only fresh updates are sent.

By combining this mechanism with SIS policies, A$^3$L-FEC optimizes sample selection, improving the freshness and reliability of updates received. This systematic approach enhances monitoring quality by prioritizing timely information. Given a fixed expected transmission rate $\tilde{\sigma}$, the selection probabilities satisfy the following relationship, used in Algorithm~\ref{alg:a3l-fec-fsfb}:

\begin{equation}
        \label{expectedthroughputforSR}
        \sum_{j=0}^{m-1}p_{j} = min(\tilde{\sigma} , AVT).
\end{equation}


Let $\Pi_{\text{SIS}}$ denote the set of all SIS policies. Our goal is to solve the following optimization problem:

\begin{equation}
        \label{optiminlimsup}
        \underset{\pi  \in \Pi _{SIS}}{\min }\underset{t \mapsto  \infty } {\limsup} \frac{1}{T} E\left [ \ \sum_{t=1}^{T} I_{\Delta(t)>AVT }   \right ]
\end{equation}


For a given expected transmission rate $\tilde{\sigma} = \min(\tilde{\sigma}, AVT)$, the A$^3$L-FEC-FSFB protocol selects the optimal probability vector $\mathbf{p}^{*}=[p_{0}^{*}(\tilde{\sigma})... \:
         p_{m-1}^{*}(\tilde{\sigma})]^{T}$  such that:

\begin{equation}
        \label{optimalp}
        p_{j}^{*}(\tilde{\sigma})=\begin{cases} 1                                            & \mbox{if } j \leq \lfloor \tilde{\sigma} \rfloor, \\
              \tilde{\sigma}-\lfloor \tilde{\sigma}\rfloor & \mbox{if } j= \lfloor \tilde{\sigma}\rfloor +1,   \\
              0                                            & \mbox{if } j > \lfloor \tilde{\sigma} \rfloor+1
        \end{cases}
\end{equation}


At time slot $t$, $p_{j}^{*}(\tilde{\sigma})$ denotes the probability of selecting a chunk from the sample generated at $t - j$. The parameter $\tilde{\sigma}$ represents the expected number of codewords transmitted per slot, each consisting of $n$ chunks.

\begin{figure}
        \centering
        \includegraphics[scale=0.5]{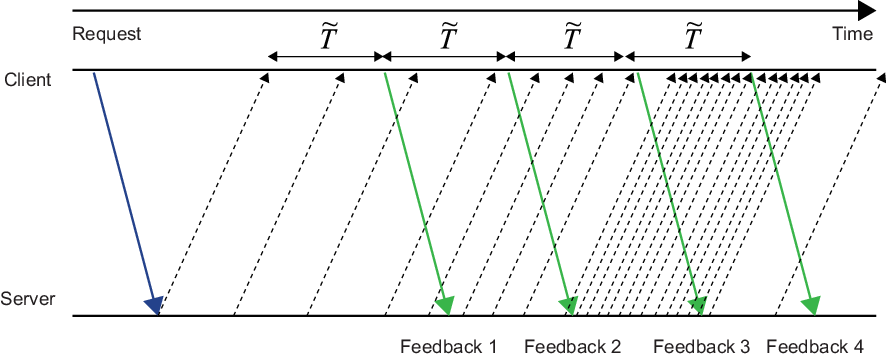}
        \caption{The signaling scheme for A$^3$L-FEC. The server acts as a transmitter and the Client acts as a receiver.}
        \label{fig:signalscheme}
\end{figure}

In the A$^3$L-FEC-FSFB protocol, the transmission rate is computed at the receiver and sent back to the transmitter every $\tilde{T}$ time slots as feedback (Fig.\ref{fig:signalscheme}). This feedback is based on the age violation metric(\ref{jkk}) and average chunk delay~(\ref{eq:delay}). Here, $\tilde{T}$ denotes the monitoring interval, and $MI$ is its index.

The age violation $AV_{MI}$ is calculated as the average number of violations during the $MI^{\text{th}}$ interval by summing $\mathbbm{1}_{{\Delta(t)> AVT}}$ over each time slot $t$ and dividing by $\tilde{T}$. The average chunk delay $\bar{W}_{MI}$ is computed as the $L_1$ norm of the delay set $\mathcal{D}_{MI}$ divided by its cardinality, representing the mean delay of successfully received chunks during that interval.

In both~(\ref{jkk}) and~(\ref{eq:delay}), $MI \in\{1,2, 3, \ldots, \lceil{\frac{T}{\tilde{T}}\rceil}\}$, and $\mathcal{D}_{MI}$ includes delays of all chunks successfully received by the end of the $MI^{\text{th}}$ monitoring interval.

\begin{equation}
        AV_{MI}=\frac{1}{\tilde{T}}\displaystyle\sum_{t=(MI-1) \times \tilde{T}}^{MI \times \tilde{T}}\mathbbm{1}_{\{ \Delta(t)> AVT \}}.
        \label{jkk}
\end{equation}

\begin{equation}
        \bar{W}_{MI} = \frac{\lVert \mathcal{D}_{MI} \rVert}{|\mathcal{D}_{MI}|}
        \label{eq:delay}
\end{equation}

\subsection{\texorpdfstring{A$^3$L-FEC}{\space}-Variable Sampling Rate Variable
    Block-length}
\label{a3lfecsystemModel_VSVB}

Similar to A$^3$L-FEC-FSFB (Section~\ref{a3lfecsystemModelFSFB}), A$^3$L-FEC-VSVB operates in a time-slotted status update system over a PEC with limited link availability. In this setup, the receiver observes a time-varying process by collecting updates from a remote transmitter, which may suffer from incomplete or lost data.

A$^3$L-FEC-VSVB follows the “\textbf{generate-at-will}” model~\cite{UpdateOrWait}, allowing the source to generate packets only when permitted by the transmission policy. This enables dynamic control of the sampling rate based on network conditions, reducing overload, delay, and age violations. Let $S$ denote the source process with samples $\{ s_{\tau} \}_{\tau \geq 0}$ generated at interval $T_s$ during each monitoring interval. Unlike the FSFB variant, both $T_s$ and block length $n$ can vary between intervals.

Each $K$-bit sample is divided into $k$ equal-sized chunks and encoded using a Maximum Distance Separable (MDS) code into $n$ chunks of size $\frac{K}{k}$ bits. The receiver can decode the sample from any $k$ of the $n$ chunks, ensuring resilience to loss. This approach is particularly suited to scenarios where retransmission is impractical, such as in deep space communications.

\begin{figure}
    \centering
    \includegraphics[scale=0.90]{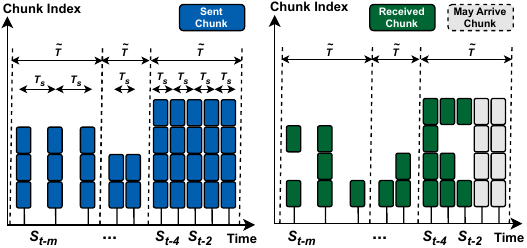}
    \caption{Chunks in the A$^3$L-FEC-VSVB protocol.}
    \label{fig:codewordsampleTX_new}


\end{figure}

As shown in Fig.~\ref{fig:codewordsampleTX_new}, the transmitter generates one sample every $T_s$ time slots, encodes it, and sends all $n$ coded chunks. The effective transmission rate is thus $n / T_s$. A key feature of A$^3$L-FEC-VSVB is its variable block length, which allows adaptation to network conditions to maintain freshness.

Coded chunks are transmitted via UDP, with each packet carrying one chunk. It is assumed that the transmitter can send $n$ packets per time slot. Let $c_{i,\tau}$ be the $i^\text{th}$ coded chunk of sample $s_\tau$, and let $\mathcal{T}_t$ be the set of packets sent at time $t$; then $c_{i,\tau} \in \mathcal{T}_t$ indicates that the chunk is sent at time $t$. The receiver stores received chunks to reconstruct the codeword, as illustrated in Fig.~\ref{fig:codewordsampleTX_new}.

\textbf{At the start of each monitoring interval}, A$^3$L-FEC-VSVB initializes
two lists to track the decode times and generation times of the decoded samples.
These lists begin with the decode time and generation time of the last sample
from the previous monitoring interval. As new samples are decoded within the
current monitoring interval, their decode and generation times are added to
these lists. Consequently, by the end of the monitoring interval, each list will
contain $C_{MI}+1$ entries, where $C_{MI}$, given by (\ref{eq:RPCounts}),
represents the total number of received chunks during the $MI^{th}$ monitoring
interval.

\begin{equation}
    C_{MI} \in \{0, 1,..., n \times [\frac{\tilde{T}}{T_s}]\}
    \label{eq:RPCounts}
\end{equation}

\textbf{At the end of each monitoring interval}, A$^3$L-FEC-VSVB performs the following five tasks:


\textbf{Task 1 – Age Violation Calculation:}
The age violation for the $MI^{\text{th}}$ monitoring interval, $AV_{MI}$, is computed using generation and decode time lists via equations~(\ref{AV-av1})–(\ref{AV-av3}), then combined in~(\ref{AV-cal}). This approach enables a more accurate measurement of AoI violations than the simplified method based on equation~(\ref{jkk}) introduced in Section~\ref{a3lfecsystemModelFSFB}.


In the following equations, let $ST_{MI}$ denote the start time of the monitoring interval, $AVT$ the age violation threshold, and $G_i$, $D_i$ the $i^\text{th}$ generation and decode times, respectively.

\begin{equation}
    \alpha =  -\Big\lvert min\Big\{ (ST_{MI} - G_{1}), max\{(ST_{MI} - G_{1} - AVT), 0\}\Big\} \Big\lvert
    \label{AV-av1}
\end{equation}

\begin{equation}
    \beta_i =  D_i - D_{i-1}
    \label{AV-av2}
\end{equation}

\begin{equation}
    \gamma_i =  AVT - (D_{i-1} - G_{i-1})
    \label{AV-av3}
\end{equation}

Therefore, the age violation is given by:
\begin{equation}
    AV_{MI} = \alpha + \sum_{i=2}^{C_{MI}} min\Bigg\{\beta_i - max\Big\{0,
    min\{\beta_i , \gamma_i\}\Big\}, \beta_i\Bigg\}.
    \label{AV-cal}
\end{equation}


\textbf{Task 2 – Chunk Delay and Packet Delivery Ratio Estimation:}
The receiver computes the average chunk delay $\bar{W}_{MI}$ and the packet delivery ratio $PDR_{MI}$ for the $MI^{\text{th}}$ monitoring interval using the following equations. Here, $W_i$ is the delay of the $i^{\text{th}}$ received chunk, $C_{MI}$ is the number of successfully received chunks during the $MI^{th}$ monitoring interval, $T_s$ is the sampling interval, and $\tilde{T}$ is the length of the monitoring interval. The denominator in~\eqref{PDR-cal} represents the expected number of transmitted packets in the interval, assuming regular sampling and codeword size $n$. These metrics enable the A$^3$L-FEC-VSVB to assess delivery efficiency and dynamically adjust the transmission rate, as detailed in Algorithm~\ref{alg:a3l-fec-vsvb}.

\begin{equation}
    \bar{W}_{MI} = \begin{cases}
        \infty                                   & \text{ if } C_{MI}= 0    \\
        \frac{\sum_{i=1}^{C_{MI}} W_{i}}{C_{MI}} & \text{ if } C_{MI}\neq 0
    \end{cases}
    \label{eq:ACD-cal}
\end{equation}

\begin{equation}
    PDR_{MI} = \frac{C_{MI}}{n \times [\frac{\tilde{T}}{T_s}]}
    \label{PDR-cal}
\end{equation}


\textbf{Task 3 – Optimal Block-Length Selection:}
Assuming delay/loss characteristics remain relatively stable between two consecutive monitoring intervals, the receiver evaluates age violations for various candidate block lengths and selects the one minimizing $AV_{MI}$. The selected $n$ is used in the next monitoring interval (see Section~\ref{ccA3L-FEC-FSFB}).


\textbf{Task 4 – Optimal Sampling Interval Calculation:}
The sample generation interval $T_s$ is updated using the selected block length $n$ and the optimal transmission rate $\sigma$:

\begin{equation}
    T_{s}=\frac{n}{\sigma}
    \label{optimal_TS}
\end{equation}


\textbf{Task 5 – Monitoring Interval Length Calculation:}
The monitoring interval length $\tilde{T}$ is determined using the age violation threshold $AVT$ to ensure zero age violations under ideal conditions, where the sample generation period equals $AVT$. The coefficient 100 guarantees sufficient chunk transmissions for reliable estimation of $PDR_{MI}$, and the coding parameter $n$ ensures alignment with the number of chunks per sample:

\begin{equation}
    \tilde{T} = [AVT \times \frac{100}{n}]
    \label{MI_duration}
    \end{equation}

\section{Congestion Control: Motivation and Overview}
\label{ccIntro}
In an ideal network with infinite resources and no delay variability, increasing the transmission rate would always improve AoI. However, in real-world settings, network congestion or path delays can degrade AoI. As a result, AoI is non-monotonic with respect to the packet injection rate, and optimizing freshness requires adaptive rate control. A$^3$L-FEC addresses this by delegating rate adaptation to the receiver, which is better positioned to assess network conditions. This receiver-driven approach enables timely adjustments based on observed chunk delays and age violations. The algorithm includes emergency response mechanisms: when the queue is empty, the rate is increased to restore update flow; when nearing overflow, the rate is reduced to mitigate congestion.

\subsection{Congestion Control Algorithm for the A$^3$L-FEC-FSFB} \label{ccA3L-FEC-FSFB}
\label{ccA3L-FEC-FSFB}
The A$^3$L-FEC-FSFB employs a receiver-driven congestion control as shown in Algorithm \ref{alg:a3l-fec-fsfb}, to dynamically adjust the transmission rate $\tilde{\sigma}$ based on real-time network conditions. The algorithm aims to balance delay and age violation by observing average packet delay ($\bar{W}_{MI}$) and age violation rate ($AV_{MI}$) within each monitoring interval. These metrics are smoothed using exponential moving averages (EMAs), enabling stable and responsive adaptation even in noisy environments.

The rate adaptation is structured around a series of decisions. When the average delay is low ($\bar{W}_{MI} < 1$) and the queue has recently been empty for multiple intervals (Empty Flag $EF \geq 2$), the algorithm anticipates rising age violations and increases the rate to refill the bottleneck. Similarly, if $\bar{W}_{MI} = \infty$ and $AV_{ema} \geq 0.9$ while $EF < 2$, this implies that insufficient chunks are in flight, and the algorithm again increases the rate.

Conversely, when both $AV_{MI}$ and $AV_{ema}$ exceed 0.9, and the average delay surpasses $AVT$, the system is likely congested. In this case, the algorithm reduces the rate to alleviate queue buildup and reduce staleness. In other scenarios, rate adjustment depends on the relative comparison between the current and EMA values of both delay and age violation.

If $AV_{MI} < AV_{ema}$ (i.e., improving condition), and delay is rising, the system probes for a higher transmission rate. If delay is falling, it suggests the previous reduction was effective, and the rate is reduced again. If $AV_{MI} > AV_{ema}$, the algorithm reduces the rate if delay is also rising (indicating congestion), or increases it if delay is falling (suggesting underutilization).

To ensure stability, the rate increase is capped at 110\% of the current rate, and rate reduction is bounded below at 20\% to prevent starvation. These safeguards are enforced in lines 21, 23, 29, and 31 of Algorithm~\ref{alg:a3l-fec-fsfb}.

Additionally, to ensure only fresh updates are transmitted, the UpdateRate function sets $\tilde{\sigma} = \min(\tilde{\sigma}, AVT)$, preventing the selection of samples older than the age violation threshold. This aligns sample freshness with system-level AoI constraints.

Parameters $\Phi$, $\Psi$, and $\Omega$ control the aggressiveness of rate changes and EMA smoothing. The Empty Flag ($EF$) tracks how many times the rate has been reduced in succession, serving as a memory for recent underutilization.

\begin{algorithm}
    \caption{A$^3$L-FEC-FSFB Congestion Control Algorithm}
    \begin{small} 
        \begin{algorithmic}[1]
            \State $AVT$ (Fixed value defined by the application) \State $\tilde{T}$
            (Fixed value defined by the application) \State $AV_{ema} \gets 0$,
            $AV_{MI} \gets 0$, $\bar{W}_{ema} \gets 0$, $\bar{W}_{MI} \gets 0$
            \State $\Phi \gets 1.5$, $\Psi \gets 0.8$, $\Omega \gets 0.8$ \State
            $MI \gets 0$, $EF \gets 0$ \State $\tilde{\sigma} \gets 2 \times n$, (n is codeword length)

            \While{true} \State $MI \gets MI + 1$ \State Transmit with rate
            $\tilde{\sigma}$ for $\tilde{T}$ timeslot \State Process($MI$)
            \If{$\bar{W}_{MI} < 1$ and $EF \geq 2$} \State$\tilde{\sigma} \gets \Phi
                \times \tilde{\sigma}$ \State$EF \gets 0$ \ElsIf{$AV_{ema} \geq
                    0.9$, $\bar{W}_{MI} == \infty$ and $EF < 2$} \State$\tilde{\sigma} \gets
                \Phi \times \tilde{\sigma}$ \ElsIf{$AV_{MI}\geq0.9$,
                $AV_{ema}\geq0.9$ and $\bar{W}_{MI}>AVT$} \State$\tilde{\sigma} \gets
                \frac{1}{\Phi} \times \tilde{\sigma} + min(0.1, \frac{1}{n})$ \Else{
            \If{$AV_{MI} \leq AV_{ema}$} \If{$\bar{W}_{MI} > \bar{W}_{ema}$}
            \State$\tilde{\sigma} \gets min(\tilde{\sigma}+(AV_{ema}-AV_{MI}),
                1.1\times\tilde{\sigma})$ \Else \State$\tilde{\sigma} \gets
                max(\tilde{\sigma}-(AV_{ema}-AV_{MI}), 0.2\times\tilde{\sigma})$
            \State$EF \gets EF+1$

            \EndIf \Else \If{$\bar{W}_{MI} > \bar{W}_{ema}$}
            \State$\tilde{\sigma} \gets max(\tilde{\sigma}-(AV_{MI}-AV_{ema}),
                0.2\times\tilde{\sigma})$ \Else \State$\tilde{\sigma} \gets
                min(\tilde{\sigma}+(AV_{MI}-AV_{ema}), 1.1\times\tilde{\sigma})$
            \EndIf

            \EndIf

            } \EndIf

            \State UpdateRate($\tilde{\sigma}, AVT$) \EndWhile \\
            \Function{UpdateRate}{$\tilde{\sigma}, AVT$} \State $\tilde{\sigma}
                \gets \min(\tilde{\sigma},AVT)$ 
            \EndFunction

            \\

            \Function{Process($MI$)}{} \State $\mathcal{D}_{MI} \gets$ Chunk's
            delays at $MI^{th}$ monitoring interval \State $ \bar{W}_{MI} \gets
                \frac{\lVert \mathcal{D}_{MI} \rVert}{|\mathcal{D}_{MI}|}$ \State
            $\bar{W}_{ema} \gets \Psi\times \bar{W}_{MI} + (1-\Psi)\times
                \bar{W}_{ema}$

            \State $AV_{MI} \gets \frac{1}{\tilde{T}}\times\sum_{t=(MI-1) \times
                    \tilde{T}}^{MI \times \tilde{T}}\mathbbm{1}_{\{ \Delta(t)>
                    AVT \}}$

            \State $AV_{ema} \gets \Omega\times AV_{MI} + (1-\Omega)\times
                AV_{ema}$

            \EndFunction

        \end{algorithmic}
    \end{small} 
    \label{alg:a3l-fec-fsfb}
\end{algorithm}

\subsection{Congestion Control Algorithm for A$^3$L-FEC-VSVB}

\label{ccA3L-FEC-VSVB}

The A$^3$L-FEC-VSVB as shown in Algorithm \ref{alg:a3l-fec-vsvb} incorporates a receiver-driven congestion control mechanism that adjusts transmission rate $\sigma$, block length $n$, sampling interval $T_s$, and the next monitoring interval ($\tilde{T}$) at the end of each monitoring interval. These adjustments are based on age violation ratio ($AV_{MI}$), average chunk delay ($\bar{W}_{MI}$), their respective exponential moving averages ($AV_{ema}$, $\bar{W}_{ema}$), the minimum observed RTT ($MinRTT$), packet delivery ratio ($PDR_{MI}$), and memory flags: Empty Flag (EF) and Decrease Flag (DF).

If no age violations are observed in the current monitoring interval, the system is considered stable and retains the last used rate ($\sigma \leftarrow \sigma_{last}$). When $\bar{W}_{MI} = \infty$, $PDR_{MI} = 0$, and $EF \geq 2$, it implies that no packets are in transit and the network is empty. In this case, the rate is aggressively increased to $\sigma = \frac{2(n + 0.05\times n)}{MinRTT}$ to refill the pipeline, including redundancy to compensate for possible packet losses.

If persistent age violations are detected and the average delay exceeds AVT, indicating congestion. The algorithm reduces $\sigma$ using a multiplicative factor $1/\Phi$ and a small additive margin $\min(0.1, 1/n)$ to ensure responsiveness without completely draining traffic. Alternatively, if age violations persist, but delay remains low and $EF \leq 2$, the system is underloaded. The rate is then increased multiplicatively to restore freshness.

In other cases, the controller compares $AV_{MI}$ to $AV_{ema}$. If the system is in a “Good” state ($AV_{MI} < AV_{ema}$), it examines $PDR_{MI}$ to guide further adjustments. When $PDR_{MI} \geq 0.9$ and $\sigma < 0.75 \times \sigma_{max}$, the system probes upward by setting $\sigma \leftarrow \sigma(1 + \frac{1}{n})$, allowing cautious exploration. If the rate is already close to $\sigma_{max}$, a dynamic increment is applied to continue probing upward while avoiding saturation. If $PDR_{MI} < 0.9$, further logic distinguishes whether delay is increasing ($\bar{W}_{MI} \geq \bar{W}_{ema}$) and no prior rate reduction occurred ($DF = False$). In that case, the rate is incremented proportionally to the gap between expected and observed age violations, scaled by the inverse of the block length: $\sigma \leftarrow \sigma + (AV_{ema} - AV_{MI}) \times \frac{1}{n}$, assuming the violation is caused by underutilization. Otherwise, if a previous reduction was applied and delay is falling, the rate is further reduced to reduce queuing and avoid triggering buffer buildup.

When in a “Bad” state ($AV_{MI} > AV_{ema}$), decisions depend on delay. If delay increases, the system is congested and $\sigma$ is decreased. If delay decreases, the age violation likely stems from under-sampling, so the rate is increased. After every update, the rate is clamped within bounds via UpdateRate, ensuring $\sigma \in [\sigma_{min}, \sigma_{max}]$ and storing $\sigma_{last}$.

This comprehensive feedback mechanism allows A$^3$L-FEC-VSVB to respond to network overload, underutilization, and dynamic delay patterns in real time, effectively managing both information freshness and network stability.

\begin{algorithm}
    \caption{A$^3$L-FEC-VSVB Congestion Control Algorithm}
    \begin{small} 
        \begin{algorithmic}[1]
            \State $AVT$ (Fixed value defined by the application) 
            \State $\tilde{T}$ (Initial value defined by the application) 
            \State $AV_{ema} \gets 0$, $AV_{MI} \gets 0$, $\bar{W}_{ema} \gets 0$, $\bar{W}_{MI} \gets 0$
            \State $\Phi \gets 1.5$, $\Psi \gets 0.8$, $\Omega \gets 0.8$ 
            \State $MI \gets 0$, $EF \gets 0$, $DF \gets False$ 
            \State $\sigma \gets \frac{2 \times (n+0.05 \times n)}{RTT_{init}}$$, \sigma_{min} \gets 0.99, \sigma_{max} \gets \frac{[4.4\times k]}{AVT} $, $\sigma_{last} \gets \sigma$

            \While{true} 
            \State $MI \gets MI + 1$ \State Transmit with rate $\sigma$ for $\tilde{T}$ timeslot 
            \State Process($MI$) \If{$AV_{MI}== 0$} 
            \State$\sigma \gets \sigma_{last}$ 
            \ElsIf{$\bar{W}_{MI} == \infty$, $EF \geq 2$ and $PDR_{MI}==0$} 
            \State$\sigma \gets \frac{2 \times (n+0.05 \times n)}{MinRTT}$ 
            \State$EF \gets 0$, $DF \gets False$
            \ElsIf{$AV_{MI} \geq 0.9$, $AV_{ema} \geq 0.9$ and $\bar{W}_{MI} \geq AVT$} 
            \State$\sigma \gets \frac{1}{\Phi} \times \sigma + min(0.1, \frac{1}{n})$ 
            \State$EF \gets EF+1$, $DF \gets True$ 
            \ElsIf{$AV_{MI} \geq 0.9$, $\bar{W}_{MI}\leq 2 \times MinRTT$, $EF \leq 2$}
            \State$\sigma \gets \Phi \times \sigma$ \State$EF \gets 0$, $DF \gets False$ 
            \Else \If{$AV_{MI} \leq AV_{ema}$} \If{$PDR_{MI} \geq 0.9$} 
            \If {$\sigma < (0.75 \times \sigma_{max})$} \State$\sigma \gets \sigma \times (1+\frac{1}{n})$ 
            \State$DF = False$ 
            \Else
            \State$\sigma \gets \sigma  + \frac{((\sigma_{max}-\sigma)+\sigma_{min}+1)}{(\sigma_{max}-\sigma_{min})}$
            \State$DF = False$ \EndIf \Else \If{$\bar{W}_{MI} \geq \bar{W}_{ema}$ and $DF == False$}\\
            ~~~~~~~~~~~~~~~~~~~~~~~~~~~ $\sigma\gets min(\sigma+(AV_{ema}-AV_{MI})\times\frac{1}{n}, 1.2\times\sigma\times\frac{1}{n})$\\
            ~~~~~~~~~~~~~~~~~~~$EF \gets 0$ 
            \Else 
            \State$\sigma\gets max(\sigma-(AV_{ema}-AV_{MI}), 0.2\times\sigma)$ 
            \State$EF \gets EF+1$, $DF = True$ 
            \EndIf 
            \EndIf

            \Else \If{$\bar{W}_{MI} > \bar{W}_{ema}$} 
            \State$\sigma \gets max(\sigma-(AV_{MI}-AV_{ema}), 0.2\times\sigma)$ 
            \State$DF \gets True$, $EF \gets 0$ 
            \Else \State$\sigma \gets min(\sigma+(AV_{MI}-AV_{ema}), 1.2\times\sigma)$ 
            \State$DF \gets False$, $EF \gets 0$ 
            \EndIf 
            \EndIf 
            \EndIf

            \State UpdateRate($\sigma_{min}, \sigma, \sigma_{max}$) 
            \EndWhile \\
            \Function{UpdateRate}{$\sigma_{min}, \sigma, \sigma_{max}$} 
            \State
            $\sigma \gets \max(\min(\sigma,\sigma_{max}),\sigma_{min})$,~~~$\sigma_{last} \gets \sigma$
            \EndFunction
            \\
            \Function{Process($MI$)}{} 
            \State $ \bar{W}_{MI} \gets$ use Eq.
            (\ref{eq:ACD-cal})
            \State $\bar{W}_{ema} \gets \Psi\times \bar{W}_{MI} + (1-\Psi)\times \bar{W}_{ema}$
            \State $AV_{MI} \gets$ use Eq. (\ref{AV-cal})
            \State $AV_{ema} \gets \Omega\times AV_{MI} + (1-\Omega)\times AV_{ema}$
            \State $MinRTT=\min(\frac{RTT}{2})$ since transmission began
            \State $n \gets$ Use task 3 in \ref{a3lfecsystemModel_VSVB}
            \State $PDR_{MI} \gets$ use Eq.\ref{PDR-cal}
            \State $T_{s} \gets$ Use Eq. \ref{optimal_TS}
            , $\tilde{T} \gets$ Use Eq. \ref{MI_duration}
            \EndFunction

        \end{algorithmic}
    \end{small} 
    \label{alg:a3l-fec-vsvb}
\end{algorithm}

\section{\texorpdfstring{A$^3$L-FEC}{ } and Theoretical Limits}
\label{TheoreticalLimitls}

\textbf{Upper Bound on Transmission Rate:} In A$^3$L-FEC, we assume UDP packets may be dropped before reaching the bottleneck with probability $P_{loss,in}$, and that the bottleneck processes packets at a rate $q_s$. To maintain queue stability, we define an upper bound on the codeword transmission rate:

\begin{equation}
\label{eq:sigamaupperbound}
\sigma^{up} = q_{s}\times \frac{1}{n\times(1-P_{loss,in})}.
\end{equation}

Here, $q_s$ is the bottleneck service rate, $n$ is the codeword length, and $P_{loss,in}$ accounts for losses before queuing. This term provides a theoretical upper bound transmission rate.

\textbf{Lower Bound on Age Violation:} To derive a lower bound on the age violation, we assume an ideal network with zero queueing delay ($q_s = 0$). For a given transmission rate $\sigma$, the decoding probability of a sample $s_{\tau}$ by time $z$ is:

\begin{align}
\label{eq:decoding}
P_{d_{s_{\tau}}}(z) = \sum_{i=0}^{n-k+1} \binom{n}{i} P_l^i (1 - P_l)^{n - i}
\end{align}

where $P_l(z - \tau)$ is the loss probability for a chunk:
\begin{align}
\label{eq:Pl}
P_l(z - \tau) = (P_{loss,c})^{z - \tau + 1}, \quad z \geq \tau
\end{align}

\textbf{Outage Probability:} Let $E_{e,t}$ be the event that the system's age is exactly $e$ at time $t$. The probability of this event is:

\begin{equation}
\label{eq:eventE}
P(E_{e,t}) = P_{d_{s_e}}(t) \cdot \prod_{j=t - e + 1}^{t} (1 - P_{d_{s_j}}(t))
\end{equation}

Accordingly, the probability that the system age exceeds a threshold $e$—known as the outage probability—captures the risk of freshness violations and serves as a bound on the worst-case performance under given transmission rates and network conditions. It is computed as:
\begin{equation}
\label{eq:outageProb}
P_{Outage}(\text{Age} > e) = 1 - \sum_{i=0}^{e} P(E_{i,t})
\end{equation}

\section{Simulation Results on MATLAB}
\label{results}

\subsection{Age Violation Evaluation: A$^3$L-FEC vs. ACP+}

We simulated the performance of A$^3$L-FEC-FSFB and compared it with ACP+ using MATLAB, replicating their core functionalities within a queuing system as described in Section~\ref{a3lfecsystemModel}. The network included a deterministic FCFS bottleneck queue with service rate $k \times 1.4706$ chunks per time slot and a buffer size of 5000 chunks. Simulations were run for $10^5$ time slots, with monitoring intervals of $10^2$ time slots, while the propagation
delay was set to 1 time slot. Packet loss probabilities before and after the bottleneck ($P_{inLoss}$, $P_{outLoss}$) were varied between 0, 0.1, and 0.2. For A$^3$L-FEC-FSFB, $k$ was set to 3 and 4, and $n$ ranged from 4 to 6.

Table~\ref{table:CompareA3lFEC-ACP-5} presents results under $AVT = 5$, with $P_{inLoss} = P_{outLoss} = 0.1$. Each entry reports the average age violation across 50 independent runs. The results show that A$^3$L-FEC-FSFB consistently achieves lower age violations than ACP+, particularly for moderate coding rates. For example, with $k=3$ and $n=4$, A$^3$L-FEC-FSFB achieves an age violation of just 0.0025, significantly outperforming ACP+'s 0.0143.

Further analysis revealed that ACP+ occasionally outperformed A$^3$L-FEC-FSFB. In those cases, A$^3$L-FEC initially transmits at a rate exceeding the bottleneck’s capacity, causing early congestion and queue buildup, which leads to temporary age violations. While the algorithm quickly reduces its rate to recover, these early violations affect its average performance.

To address this, we propose two enhancements: (1) a slow-start mechanism that gradually ramps up the rate based on feedback, and (2) adaptive monitoring intervals that enable quicker feedback and more responsive rate adjustments. These ideas motivated the development of A$^3$L-FEC-VSVB.

\begin{table}[htbp]
\begin{center}
\caption{A$^{3}$L-FEC-FSFB vs. ACP+}
\begin{tabular}{ |c|c|c|c|} 
\hline
\multicolumn{4}{|c|}{\textbf{A$^{3}$L-FEC-FSFB}} \\
\hline
    & n = 4 & n = 5 & n = 6  \\
\hline
 k = 3   & AV = 0.0025 & AV = 0.0095 & AV = 0.0200  \\
\hline
 k = 4   & - & AV = 0.0031 & AV = 0.0101  \\
\hline
\hline
\multicolumn{4}{|c|}{\textbf{ACP+}} \\
\hline
\multicolumn{4}{|c|}{AV = 0.0143} \\
\hline

\end{tabular}
\label{table:CompareA3lFEC-ACP-5}
\end{center}
\end{table}

\subsection{Effect of Coding Rate on A$^3$L-FEC-FSFB} \label{resultsOnPc2}

To evaluate the impact of coding rate on the performance of A$^3$L-FEC-FSFB, we conducted simulations using the setup described in Section~\ref{a3lfecsystemModel}. Transmissions occurred over a deterministic FCFS queue with a service rate of $k \times 1.4706$ chunks per time slot and a buffer size of 5000 chunks. Each run lasted $10^5$ time slots, with monitoring intervals of $10^2$ time slots. Simulations were repeated 50 times for statistical consistency, and parameters are summarized in Table~\ref{tab:MatlabSimulationParametersCodingRate}.

\begin{table}
\begin{center}
\caption {MATLAB Simulation Parameters for Evaluating the Effect of Coding Rate on the Performance of A$^3$L-FEC-FSFB}
\normalsize

\begin{tabular}{ |c||c| }
    \hline
    \textbf{Age Violation Threshold} & 2, 5 [time slot] \\
    \hline
    \textbf{Bottleneck Queue Size} & $5000$ [chunk]   \\
    \hline
    \textbf{Queue service rate} &  $k \times 1.4706$  $[\frac{packets}{time slot}]$  \\
    \hline
   \textbf{$P_{inLoss}$} & $ 0, 0.1, 0.2$ \\
    \hline
   \textbf{ $P_{outLoss}$}& $0, 0.1, 0.2$ \\
    \hline
         \textbf{Propagation Delay}  & $1$ [time slot] \\
    \hline
        \textbf{MI Duration} & $10^2$ [time slot]\\
    \hline
        \textbf{Simulation Duration} & $10^5$ [time slot] \\
    \hline
       \textbf{ k for A$^3$L-FEC-FSFB} & 2, 3, 4 \\
    \hline
        \textbf{n for A$^3$L-FEC-FSFB} & 2, 3, 4, 5, 6, 7, 8, 9 \\
    \hline
    
\end{tabular}
\label{tab:MatlabSimulationParametersCodingRate}
\end{center}
\end{table}

\begin{figure} \centering \includegraphics[scale=0.53]{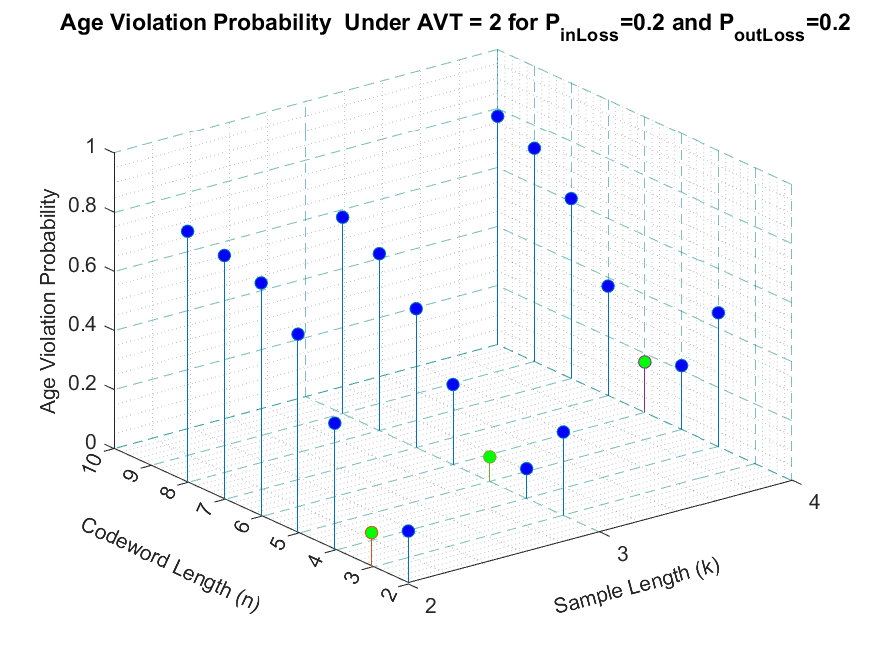} \caption{Age violation under different coding rates, ($AVT = 2$).} \label{fig:a2p02p02} \end{figure}

\begin{figure} \centering \includegraphics[scale=0.53]{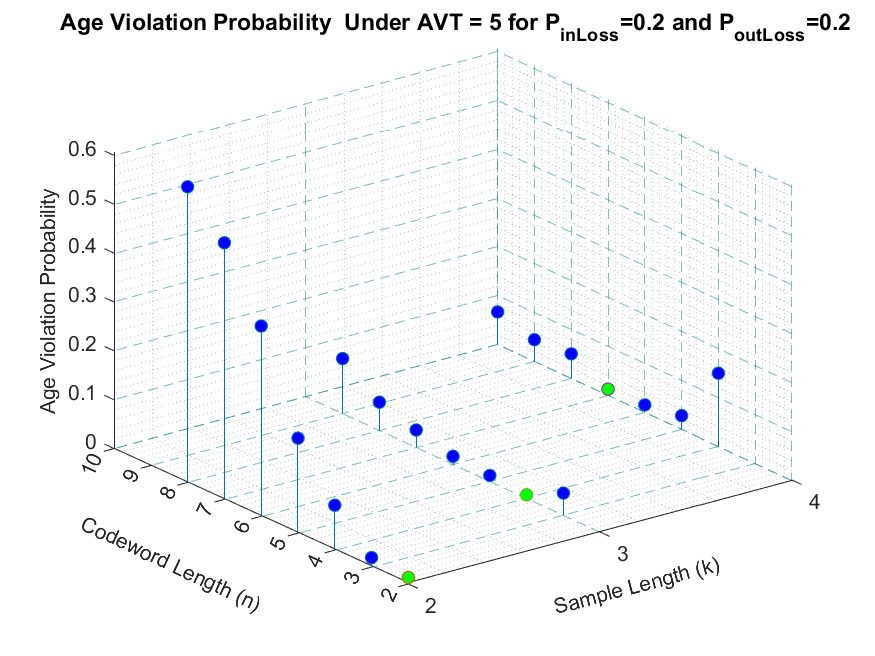} \caption{Age violation under different coding rates, ($AVT = 5$).} \label{fig:a5p02p02} \end{figure}

We considered two age violation thresholds, $AVT = 2$ and $AVT = 5$, under different packet loss settings ($P_{inLoss}$, $P_{outLoss} \in {0, 0.1, 0.2}$). Coding parameters $k$ and $n$ were varied across a wide range. Figures~\ref{fig:a2p02p02} and~\ref{fig:a5p02p02} show sample results for $AVT = 2$ and $AVT = 5$, both with $P_{inLoss} = P_{outLoss} = 0.2$. The results highlight a clear trade-off between redundancy and delay: increasing $n$ improves decoding reliability but adds network load and queuing delay. The optimal coding rate—marked by the green point—balances these effects and significantly reduces age violations.

While redundancy is essential in lossy networks, excessive coding inflates queue sizes and delays. Thus, choosing an appropriate coding rate is critical, even in scenarios with low average packet loss. These findings informed the development of the adaptive A$^3$L-FEC-VSVB scheme, introduced in Section~\ref{a3lfecsystemModel_VSVB}, which dynamically adjusts coding parameters.

\section{AoI Evaluation of TCP on ns-3}
\label{ns3}

\begin{figure}
  \centering
  \includegraphics[scale=0.5]{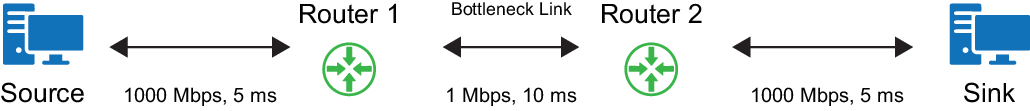}
  \caption{End-to-end network topology used for simulation.}
  \label{fig:topology1}
\end{figure}

We used the ns-3 simulator to evaluate the AoI performance of several TCP congestion control algorithms, including TCP BBR, BIC, Cubic, Ledbat, and NewReno. The goal was to assess their effectiveness in maintaining information freshness under congestion. The simulated topology (Fig.~\ref{fig:topology1}) consists of one source, two routers, and one sink. A 10 Mbps bottleneck link between the routers and DropTail queuing (100-packet limit) introduced controlled congestion, while all other links had 1000 Mbps capacity. Propagation delays were set to 10 ms for the bottleneck and 5 ms elsewhere. 

As shown in Table~\ref{tab:tcp-aoi}, TCP BBR outperformed other variants, achieving the lowest average AoI. This improvement stems from BBR’s conservative rate control and minimal queuing behavior, which reduce delay and packet loss. Unlike loss-based TCPs that overfill buffers, BBR injects packets based on bandwidth estimation, keeping queues short and ensuring timely updates. These characteristics make BBR an ideal candidate as a benchmark for comparison with A$^3$L-FEC in subsequent evaluations. 


\begin{table}[htbp]
\centering
\caption{Average AoI for TCP Variants}
\label{tab:tcp-aoi}
\begin{tabular}{|l|c|}
\hline
\textbf{TCP Variant} & \textbf{Average AoI ($\mu$s)} \\
\hline
TCP BBR       & 1.1 \\
TCP NewReno   & 2.5 \\
TCP Ledbat    & 4.05 \\
TCP BIC       & 4.7 \\
TCP Cubic     & 4.8 \\
\hline
\end{tabular}
\end{table}

\section{Emulation Results on Mininet-WiFi} 
\label{A3LFEC-FSFB-BBR-mininet}

Motivated by the strong age performance of BBR identified in Section~\ref{ns3}, we conducted a series of emulation experiments to compare it against the proposed A$^3$L-FEC protocol. The experiments were carried out on Mininet-WiFi, using a custom end-to-end topology to simulate realistic wireless network conditions. Each scenario was repeated three times, and results were averaged to ensure consistency and minimize randomness.

To implement A$^3$L-FEC-VSVB, we developed C++ client-server applications running over UDP, with the congestion control logic implemented at the application layer. This setup enables end-to-end age-aware decisions and provides flexibility for application-specific optimizations.

\begin{figure}
\centering
\includegraphics[scale=0.8]{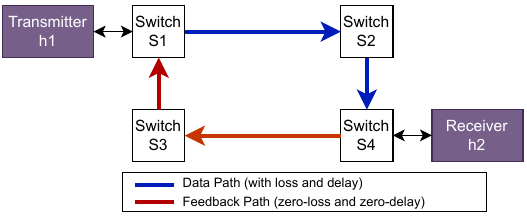}
\caption{End-to-end network topology used for emulation.}
\label{fig:Emulatormodel}
\end{figure}

The emulated topology (Fig.~\ref{fig:Emulatormodel}) includes one source node (h1), one sink node (h2), and four switches. The bottleneck lies between switches s1 and s4 via s2, configured with 1 Mbps bandwidth and a 1000-packet DropTail buffer. All other links offer 1000 Mbps capacity. The forward propagation delay from h1 to h2 is 30 ms; the return path has zero delay. A 5\% packet loss rate is applied only on the bottleneck link.


Data flows from h1 to h2 through switches s1–s2–s4 using FCFS queuing, while feedback from the receiver is sent back via s4–s3–s1. The A$^3$L-FEC-VSVB protocol dynamically adjusts its sampling interval and coding rate based on real-time feedback. In contrast, TCP-BBR was evaluated with fixed sampling rates of 100, 200, 300, and 600 ms.





\subsection{Age Violation Evaluation: A$^3$L-FEC-VSVB vs. TCP-BBR}
\label{results:AV}

Figures~\ref{fig:AV_06} and~\ref{fig:AV_015} show the age violation performance of A$^3$L-FEC-VSVB and TCP-BBR when transmitting 500 samples under age thresholds of 600 ms and 150 ms. A$^3$L-FEC-VSVB consistently outperforms TCP-BBR across all tested configurations, demonstrating its effectiveness in maintaining information freshness.

TCP-BBR suffers from increased age violations when paired with a fixed 100 ms sampling interval. This is primarily due to the mismatch between sample generation and network service capacity, which leads to queuing delays at the transmitter. Although TCP-BBR performs well at lower sampling rates, its fixed configuration hinders adaptation to changing conditions.

By contrast, A$^3$L-FEC-VSVB dynamically adjusts its sampling period and block length in response to network feedback, effectively minimizing age violations. Even when TCP-BBR performs reasonably well (e.g., at 200 ms sampling), A$^3$L-FEC-VSVB maintains superior performance due to its adaptive, age-aware control strategy.

\subsection{AoI Evaluation: A$^3$L-FEC-VSVB vs. TCP-BBR}

\begin{figure*}[t]
        \centering
        \begin{subfigure}[b]{0.49\textwidth}
            \includegraphics[width=\textwidth]{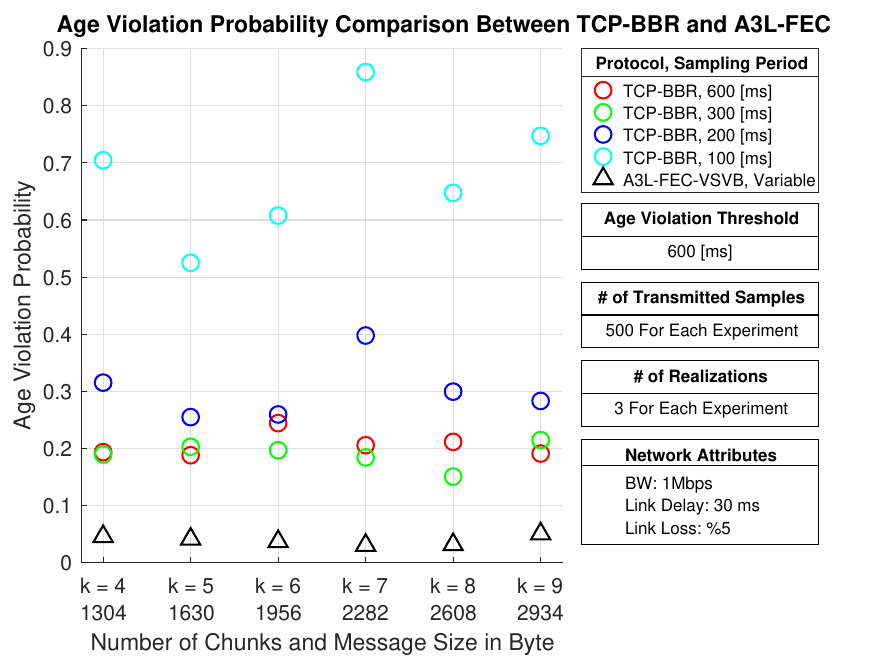}
            \caption{}
            \label{fig:AV_06}
        \end{subfigure}
        \begin{subfigure}[b]{0.49\textwidth}
            \includegraphics[width=\textwidth]{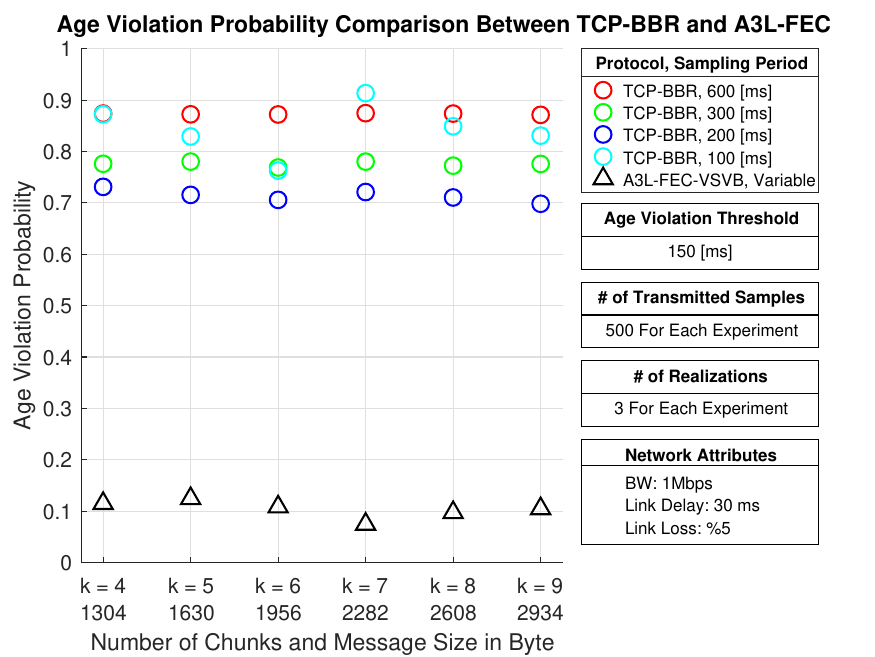}
            \caption{}
            \label{fig:AV_015}
        \end{subfigure}
        \vspace{0.5em}

        \begin{subfigure}[b]{0.49\textwidth}
            \includegraphics[width=\textwidth]{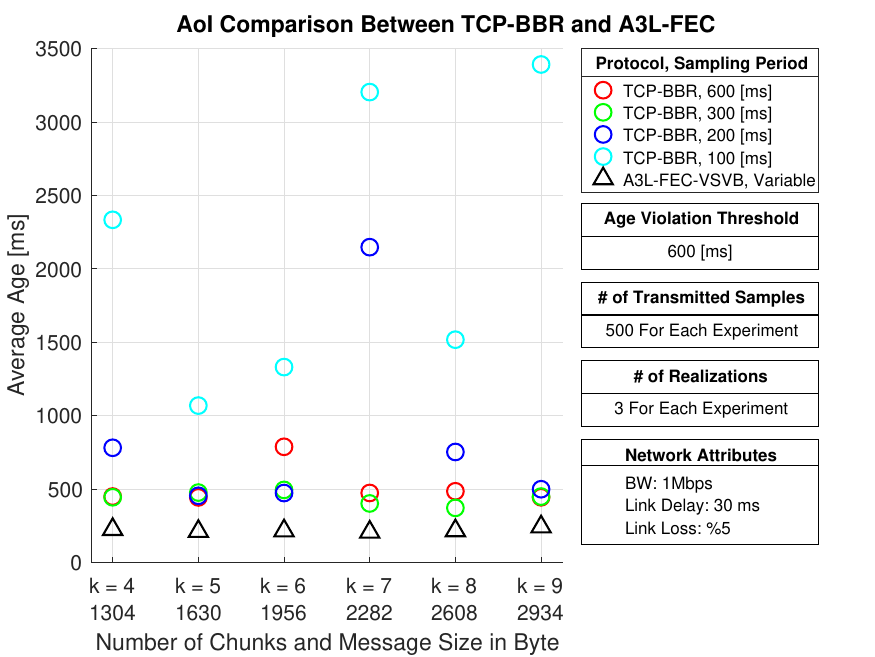}
            \caption{}
            \label{fig:EmulatorAoI600}
        \end{subfigure}
        \begin{subfigure}[b]{0.49\textwidth}
            \includegraphics[width=\textwidth]{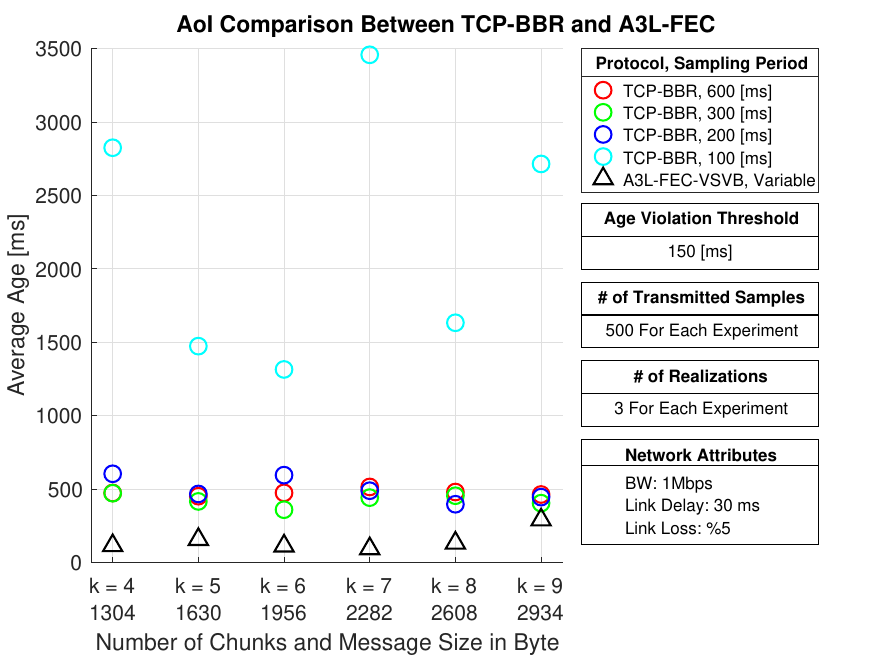}
            \caption{}
            \label{fig:EmulatorAoI150}
        \end{subfigure}
        \vspace{0.5em}

        \begin{subfigure}[b]{0.49\textwidth}
            \includegraphics[width=\textwidth]{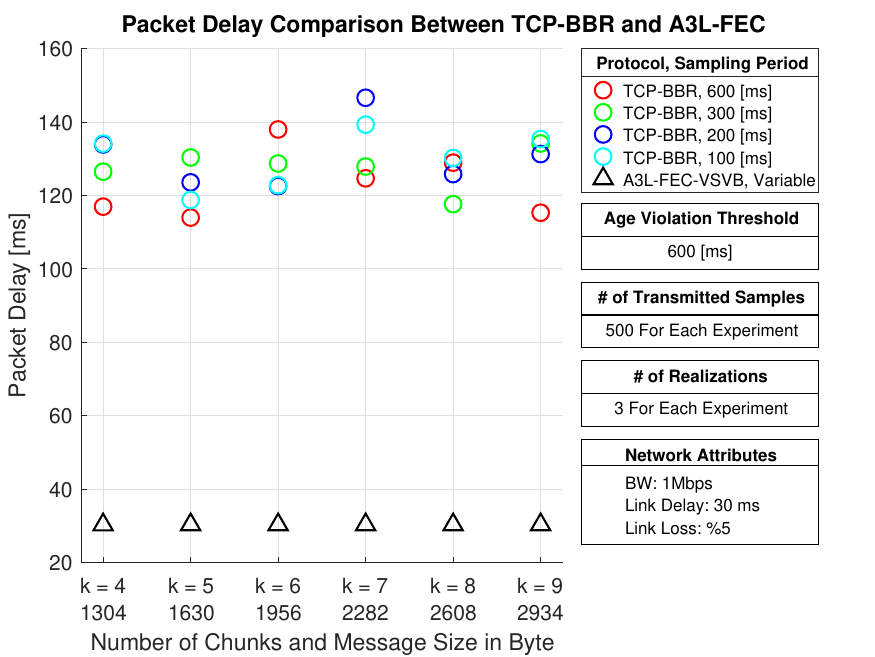}
            \caption{}
            \label{fig:EmulatorDelay600}
        \end{subfigure}
        \begin{subfigure}[b]{0.49\textwidth}
            \includegraphics[width=\textwidth]{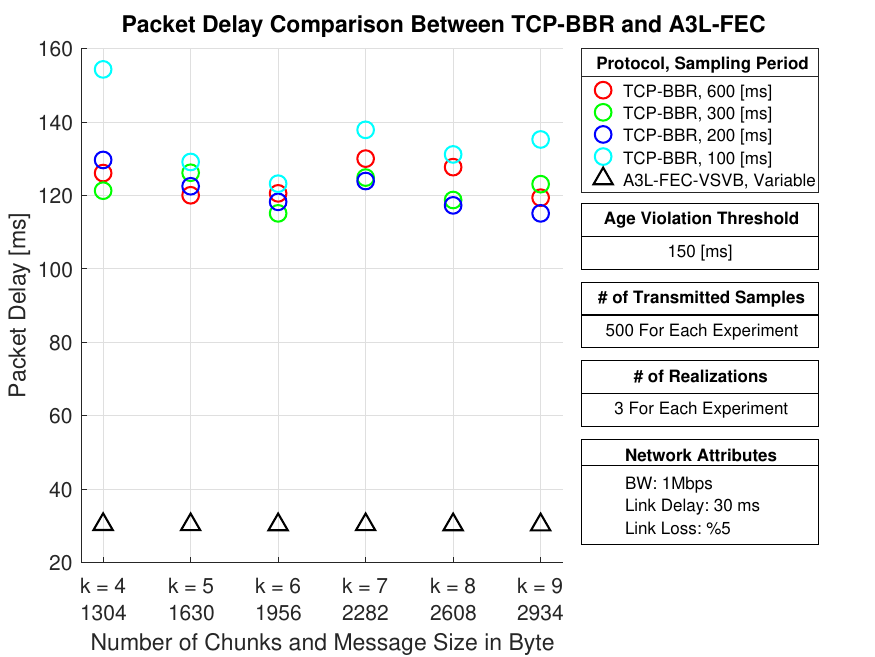}
            \caption{}
            \label{fig:EmulatorDelay150}
        \end{subfigure}

        \caption{Overall caption for the figure.}
        \label{fig:foobar}
    \end{figure*}

Figures \ref{fig:EmulatorAoI600} and \ref{fig:EmulatorAoI150} demonstrate that A$^3$L-FEC-VSVB consistently achieves lower AoI compared to TCP-BBR across all tested scenarios. This advantage stems from the fundamental differences in their transmission strategies. TCP-BBR buffers generated samples and transmits them as bandwidth allows.

\FloatBarrier
\FloatBarrier

Newly generated samples may queue behind older ones, leading to increased age and occasional violations. In contrast, A$^3$L-FEC-VSVB employs a \textit{\textbf{generate-at-will}} policy combined with a \textit{\textbf{fire-and-forget}} mechanism, transmitting each sample immediately without queuing.

Moreover, A$^3$L-FEC operates over UDP, eliminating the need for connection establishment or retransmission, further reducing delays. This streamlined, application-layer design enables A$^3$L-FEC to maintain fresher data at the receiver, making it more suitable for age-sensitive applications.

\subsection{Packet Delay Evaluation: A$^3$L-FEC-VSVB vs. TCP-BBR} \label{results:delay}

Figures \ref{fig:EmulatorDelay600} and \ref{fig:EmulatorDelay150} present the average packet delay for transmitting 500 samples under varying conditions. The results show that A$^3$L-FEC-VSVB consistently maintains a lower and more stable delay—approximately 30 ms—which closely matches the propagation delay of the network. In contrast, TCP-BBR exhibits an average delay of about 130 ms due to its reliance on connection setup and retransmission mechanisms.

This difference stems from the core design philosophies of the two protocols. A$^3$L-FEC, operating over UDP, avoids retransmissions and connection handshakes, employing a fire-and-forget strategy that ensures immediate transmission without buffering. TCP-BBR, on the other hand, uses a three-way handshake to establish connections and retransmits lost packets, introducing additional latency.

By eliminating retransmissions and startup overhead, A$^3$L-FEC achieves faster end-to-end delivery, making it more suitable for latency-sensitive applications. Its age-aware congestion control further helps maintain low queuing delays, ensuring timely packet delivery even under dynamic network conditions.

\subsection{Effect of Chunk Number on A$^3$L-FEC-VSVB}
\label{results:codingrate}

\begin{figure}
\centering
\includegraphics[scale=0.6]{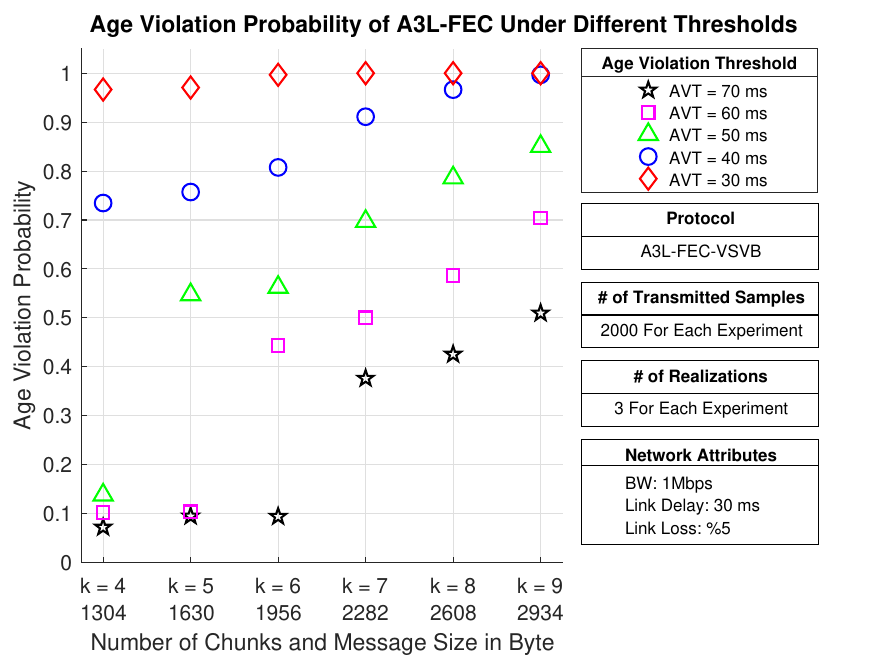}
\caption{Effect of Sample Chunk Number and Age Violation Threshold on the Performance of A$^3$L-FEC-VSVB}
\label{fig:EmulatorchunkNumber}
\end{figure}

To evaluate the impact of chunk number on the performance of A$^3$L-FEC-VSVB, we emulated the transmission of 2000 samples over Mininet-WiFi, varying the age violation threshold across 70, 60, 50, and 40 ms. The results, shown in Fig.~\ref{fig:EmulatorchunkNumber}, illustrate how both chunk size and threshold settings influence age violation behavior.

As the number of chunks per sample increases, the probability of age violation also rises. Although only $k$ out of $n$ chunks are required for decoding, higher $n$ values lead to longer transmissions, elevated network load, and increased buffering, all of which contribute to delay and congestion. For higher thresholds (70 ms and 60 ms), A$^3$L-FEC-VSVB maintains low age violation rates, even with moderate chunk counts. However, performance degrades significantly for lower thresholds (50 ms, 40 ms), particularly at larger coding rates.

This highlights a trade-off: larger chunk sizes offer greater redundancy and reliability but at the cost of increased transmission time and queuing delay. Smaller chunk sizes help maintain freshness but may compromise decoding success under loss.

\section{\texorpdfstring{A$^3$L-FEC}{\space} in Multiserver Scenarios} 
\label{A3L-FEC-Multiserver}

To reflect practical network environments where multiple transmitters and receivers share limited resources, we extend the A$^3$L-FEC system model to a multi-transmitter, multi-receiver setting. In this scenario, all flows traverse a common bottleneck, introducing contention and the need for coordinated congestion control. The goal is to allocate transmission rates across competing sources while minimizing age violations.

The congestion control manages the total system rate, denoted by $\sigma_{Total}$, which is calculated based on observed network conditions. This total rate is then distributed among the transmitters proportionally, taking into account each transmitter's current age violation level. Transmitters with higher age violations are assigned higher rates to accelerate fresh data delivery, while those with lower violations receive reduced rates. This ensures fairness and responsiveness across the system.

Rate allocation is governed by the rule in Equation~\eqref{eq:multiserver-sigma}, which adjusts each transmitter $i$’s rate based on its previous rate $\sigma_i^{Old}$, its own age violation $AV_{MI,i}$, the system-wide average age violation $AV_{MIav}$, and the ratio of updated total rates between intervals:

\begin{equation} \sigma_{i} = \left ( \sigma_i^{Old} + (AV_{MI,i} - AV_{MIav}) \right )\times \left ( \frac{\sigma_{Total}}{\sigma_{Total}^{Old}} \right ) \label{eq:multiserver-sigma} \end{equation}

Here, the term $(AV_{MI,i} - AV_{MIav})$ identifies whether transmitter $i$ is performing above or below the system average in terms of freshness. If its age violation is higher than average, the term is positive and leads to a rate increase. If it is lower, the transmitter’s rate is reduced, freeing resources for other flows in need. The scaling factor $\left( \frac{\sigma_{Total}}{\sigma_{Total}^{Old}} \right)$ ensures that changes in total system capacity are fairly distributed across all transmitters.

This adaptive strategy enables coordinated rate control while keeping the overall transmission load within system limits. By continuously adjusting each transmitter’s rate based on age violation trends, the algorithm promotes balanced freshness across flows, mitigates congestion, and maximizes system efficiency.

\section{Conclusions}
\label{Conclusions}

This paper introduced A$^3$L-FEC, a novel age-aware flow control protocol that improves data freshness by minimizing the rate of age violations at the application layer. Built on UDP and enhanced with packet-level FEC, A$^3$L-FEC avoids retransmissions while ensuring timely delivery. Two variants, FSFB and VSVB, support flexible sampling and adaptive block-length selection.

Both simulation and emulation results demonstrate that A$^3$L-FEC consistently outperforms protocols like TCP-BBR and ACP+, achieving lower age violations, reduced delays, and better adaptability in lossy, time-sensitive networks.

A multiserver extension was also proposed to coordinate rate control across competing flows. Future work will refine this extension and explore real-world deployment in IoT systems, mobile networks, and delay-tolerant applications—including deep space missions and autonomous systems, with a focus on scalability, robustness, and domain-specific performance.

\section*{Acknowledgment}
This study was supported by Scientific and Technological Research Council of Turkey (TUBITAK), Grant 22AG019. We are also grateful to Dr. Baran Tan Bacınoğlu and Dr. Mahdi Shakiba Herfeh for their contributions to analysis and insights.  

\bibliographystyle{IEEEtran}
\bibliography{IEEEabrv,Bibliography}

\begin{thebibliography}{10}
\providecommand{\url}[1]{#1}
\csname url@rmstyle\endcsname
\providecommand{\newblock}{\relax}
\providecommand{\bibinfo}[2]{#2}
\providecommand\BIBentrySTDinterwordspacing{\spaceskip=0pt\relax}
\providecommand\BIBentryALTinterwordstretchfactor{4}
\providecommand\BIBentryALTinterwordspacing{\spaceskip=\fontdimen2\font plus
\BIBentryALTinterwordstretchfactor\fontdimen3\font minus
  \fontdimen4\font\relax}
\providecommand\BIBforeignlanguage[2]{{%
\expandafter\ifx\csname l@#1\endcsname\relax
\typeout{** WARNING: IEEEtran.bst: No hyphenation pattern has been}%
\typeout{** loaded for the language `#1'. Using the pattern for}%
\typeout{** the default language instead.}%
\else
\language=\csname l@#1\endcsname
\fi
#2}}

\bibitem{2011KaulGruteserRaiKenney}
S.~Kaul, M.~Gruteser, V.~Rai, and J.~Kenney, ``Minimizing age of information in
  vehicular networks,'' in \emph{2011 8th Annual IEEE Communications Society
  Conference on Sensor, Mesh and Ad Hoc Communications and Networks}, 2011, pp.
  350--358.

\bibitem{10634462}
E.~Markova, V.~E. Manaeva, E.~Zhbankova, D.~Moltchanov, P.~Balabanov,
  Y.~Koucheryavy, and Y.~Gaidamaka, ``Performance-utilization trade-offs for
  state update services in 5g nr systems,'' \emph{IEEE Access}, vol.~12, 2024.

\bibitem{10415249}
J.~Wang, L.~Bai, Z.~Fang, R.~Han, J.~Wang, and J.~Choi, ``Age of information
  based urllc transmission for uavs on pylon turn,'' \emph{IEEE Transactions on
  Vehicular Technology}, vol.~73, no.~6, 2024.

\bibitem{10355071}
Y.~Chen, H.~Lu, L.~Qin, Y.~Deng, and A.~Nallanathan, ``When xurllc meets noma:
  A stochastic network calculus perspective,'' \emph{IEEE Communications
  Magazine}, vol.~62, no.~6, 2024.

\bibitem{10663282}
F.~Khodakhah, A.~Mahmood, c.~Stefanovic, H.~Farag, P.~Österberg, and
  M.~Gidlund, ``Balancing aoi and rate for mission-critical and embb
  coexistence with puncturing, noma, and rsma in cellular uplink,'' \emph{IEEE
  Transactions on Vehicular Technology}, vol.~74, no.~1, 2025.

\bibitem{Kleinrock2018InternetCC}
L.~Kleinrock, ``Internet congestion control using the power metric: Keep the
  pipe just full, but no fuller,'' \emph{Ad Hoc Networks}, vol.~80, 2018.

\bibitem{Turkovic2019FiftySO}
B.~Turkovic, F.~A. Kuipers, and S.~Uhlig, ``Fifty shades of congestion control:
  A performance and interactions evaluation,'' \emph{ArXiv}, vol.
  abs/1903.03852, 2019.

\bibitem{survey2}
R.~D. Yates, Y.~Sun, D.~R. Brown, S.~K. Kaul, E.~H. Modiano, and S.~Ulukus,
  ``Age of information: An introduction and survey,'' \emph{IEEE Journal on
  Selected Areas in Communications}, vol.~39, 2021.

\bibitem{Uysal_Kaya_Baghaee_Beytur_2023}
E.~Uysal, O.~Kaya, S.~Baghaee, and H.~B. Beytur, \emph{Age of Information in
  Practice}.\hskip 1em plus 0.5em minus 0.4em\relax Cambridge University Press,
  2023, p. 297–326.

\bibitem{core-aoi}
C.~Kam, S.~Kompella, and A.~Ephremides, ``Experimental evaluation of the age of
  information via emulation,'' in \emph{MILCOM 2015 - 2015 IEEE Military
  Communications Conference}, Oct 2015, pp. 1070--1075.

\bibitem{canberk2018}
C.~Sönmez, S.~Baghaee, A.~Ergişi, and E.~Uysal-Biyikoglu,
  ``\text{Age-of-Information} in practice: Status age measured over
  \text{TCP/IP} connections through \text{WiFi, Ethernet and LTE},'' in
  \emph{2018 IEEE International Black Sea Conference on Communications and
  Networking (BlackSeaCom)}, June 2018, pp. 1--5.

\bibitem{HasanRealLife}
H.~B. {Beytur}, S.~{Baghaee}, and E.~{Uysal}, ``Measuring age of information on
  real-life connections,'' in \emph{2019 27th Signal Processing and
  Communications Applications Conference (SIU)}, 2019, pp. 1--4.

\bibitem{HasanSmartIoT}
H.~B. Beytur, S.~Baghaee, and E.~Uysal, ``Towards aoi-aware smart iot
  systems,'' in \emph{2020 International Conference on Computing, Networking
  and Communications (ICNC)}, 2020, pp. 353--357.

\bibitem{EgemenRL}
E.~Sert, C.~Sönmez, S.~Baghaee, and E.~Uysal-Biyikoglu, ``Optimizing age of
  information on real-life \text{TCP/IP} connections through reinforcement
  learning,'' in \emph{2018 26th Signal Processing and Communications
  Applications Conference (SIU)}, May 2018, pp. 1--4.

\bibitem{10032102}
H.~Wang, X.~Xie, and J.~Yang, ``Optimizing average age of information in
  industrial iot systems under delay constraint,'' \emph{IEEE Transactions on
  Industrial Informatics}, vol.~19, no.~10, pp. 10\,244--10\,253, 2023.

\bibitem{kadota2020wifresh}
I.~Kadota, M.~S. Rahman, and E.~Modiano, ``\text{WiFresh}: Age-of-information
  from theory to implementation,'' 2020.

\bibitem{10228860}
V.~Tripathi, I.~Kadota, E.~Tal, M.~S. Rahman, A.~Warren, S.~Karaman, and
  E.~Modiano, ``Wiswarm: Age-of-information-based wireless networking for
  collaborative teams of uavs,'' in \emph{IEEE INFOCOM 2023 - IEEE Conference
  on Computer Communications}, 2023, pp. 1--10.

\bibitem{ACP2018}
T.~Shreedhar, S.~K. Kaul, and R.~D. Yates, ``Acp: Age control protocol for
  minimizing age of information over the internet,'' in \emph{Proceedings of
  the 24th Annual International Conference on Mobile Computing and Networking},
  ser. MobiCom '18.\hskip 1em plus 0.5em minus 0.4em\relax New York, NY, USA:
  ACM, 2018.

\bibitem{10483026}
------, ``Acp $+$ : An age control protocol for the internet,'' \emph{IEEE/ACM
  Transactions on Networking}, pp. 1--16, 2024.

\bibitem{ACPUmut}
U.~Guloglu, S.~Baghaee, and E.~Uysal, ``Evaluation of age control protocol
  (acp) and acp+ on esp32,'' in \emph{2021 17th International Symposium on
  Wireless Communication Systems (ISWCS)}, 2021, pp. 1--6.

\bibitem{costa2020robust}
M.~Costa, Y.~Sagduyu, T.~Erpek, and M.~Médard, ``Robust improvement of the age
  of information by adaptive packet coding,'' 2020.

\bibitem{10601159}
S.~Baghaee and E.~Uysal, ``A3l-fec-fsfb: Age-aware application layer forward
  error correction with fixed sampling rate and fixed block-length,'' in
  \emph{2024 32nd Signal Processing and Communications Applications Conference
  (SIU)}, 2024, pp. 1--4.

\bibitem{UpdateOrWait}
Y.~Sun, E.~Uysal-Biyikoglu, R.~D. Yates, C.~E. Koksal, and N.~B. Shroff,
  ``Update or wait: How to keep your data fresh,'' \emph{IEEE Transactions on
  Information Theory}, vol.~63, no.~11, pp. 7492--7508, Nov 2017.

\end{thebibliography}

\begin{IEEEbiography}
[{\includegraphics[width=1in,height=1.25in,clip,keepaspectratio]{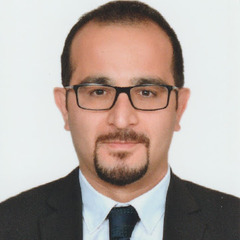}}]{Sajjad Baghaee} (Member, IEEE) is a Senior IoT Engineer and Postdoctoral Researcher at Middle East Technical University (METU), Ankara, Turkey. With over 14 years of experience, he specializes in information freshness, age-aware flow control, and goal-oriented communications. He has led and contributed to several national and industry-funded R\&D projects in smart energy systems, LoRa/LoRaWAN-based monitoring, and delay-tolerant networks. Dr. Baghaee holds a Ph.D. in Electrical and Electronics Engineering from METU, where his research focused on Age of Information-based flow control. His work spans embedded systems, protocol design (MQTT, LoRaWAN, TCP/UDP), and the application of machine learning in IoT systems. He has authored over 25 publications and a patent, and his current research interests include age-sensitive communication protocols for IoT, real-time systems, and 6G technologies.

\end{IEEEbiography}
\begin{IEEEbiography}[{\includegraphics[width=1in,height=1.25in,clip,keepaspectratio]{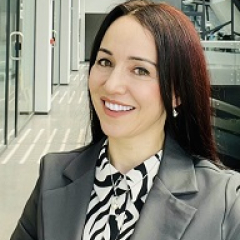}}]{Elif Uysal} (Fellow, IEEE) is a Professor in the Department of Electrical and Electronics Engineering at Middle East Technical University (METU), Ankara, Turkey. She received her Ph.D. in Electrical Engineering from Stanford University in 2003, the S.M. degree in EECS from MIT in 1999, and the B.S. degree from METU in 1997. Prior to joining METU in 2006, she held faculty positions at MIT and Ohio State University.

Her research lies at the intersection of communication and networking theory, with a focus on energy-efficient and low-latency wireless systems. She was named IEEE Fellow in 2022 for her pioneering contributions to this field. Prof. Uysal is also a Fellow of the Asia-Pacific Artificial Intelligence Association and the Artificial Intelligence Industry Alliance. She has received numerous awards, including the TÜBİTAK Pioneer Researcher Grant, Science Academy Young Scientist Award (2014), IBM Faculty Award (2010), and both NSF and TÜBİTAK career awards. Prof. Uysal is the founder of FRESHDATA Technologies (2022) and serves as Chair of the Executive Board and Board of Trustees of the METU Parlar Foundation. She has served on the editorial boards of the IEEE/ACM Transactions on Networking and IEEE Transactions on Wireless Communications, and as lead guest editor for the IEEE Journal on Selected Areas in Information Theory. Her service includes leadership roles in numerous IEEE conferences, including ICC, INFOCOM, WiOpt, SIU, and ISIT.

\end{IEEEbiography}
\end{document}